\newif\ifsubmode
\submodefalse

\newif\ifprintfig
\printfigtrue

\newif\ifemulate
\emulatetrue

\ifsubmode
  \documentclass[12pt,preprint]{aastex}
  \received{}
  \accepted{}
  \journalid{}{}
  \articleid{}{}
\else
   \documentclass{emulateapj}
   \submitted{{\it Submitted for publication in AJ}}
\fi

\shortauthors{Willman, et al.}
\shorttitle{New Milky Way Companion}

\def\lesssim{\mathrel{\hbox{\rlap{\hbox{\lower4pt\hbox{$\sim$}}}\hbox{$<$}}}}
\def\gtrsim{\mathrel{\hbox{\rlap{\hbox{\lower4pt\hbox{$\sim$}}}\hbox{$>$}}}}

\begin{document}

\title{A New Milky Way Companion: Unusual Globular Cluster or Extreme Dwarf Satellite?}

\author{Beth Willman\altaffilmark{1}, Michael R. Blanton\altaffilmark{1}, Andrew A. West\altaffilmark{2}, Julianne J. Dalcanton\altaffilmark{2,3}, David W. Hogg\altaffilmark{1}, Donald P. Schneider\altaffilmark{4}, Nicholas Wherry\altaffilmark{1}, Brian Yanny\altaffilmark{5}, Jon Brinkmann\altaffilmark{6} }


\altaffiltext{1}{New York University, Center for Cosmology and Particle Physics,
4 Washington Place, New York, NY 10003}
 
\altaffiltext{2}{Department of Astronomy, University of Washington, Box 351580, Seattle, WA, 98195}

\altaffiltext{3}{Alfred P. Sloan Research Fellow}

\altaffiltext{4}{Department of Astronomy and Astrophysics, Penn State University, University Park, PA 16802}

\altaffiltext{5}{Fermi National Accelerator Laboratory, PO Box 500, Batavia, IL 60510}

\altaffiltext{6}{Apache Point Observatory, PO Box 59, Sunspot, NM 88349}


\ifsubmode\else
  \ifemulate\else
     \clearpage
  \fi
\fi


\ifsubmode\else
  \ifemulate\else
     \baselineskip=14pt
  \fi
\fi

\begin{abstract}
We report the discovery of SDSSJ1049+5103, an overdensity of resolved
blue stars at ($\alpha_{2000}$, $\delta_{2000}$) = (162.343, 51.051).
This object appears to be an old, metal-poor stellar system at a
distance of 45$\pm 10$ kpc, with a half-light radius of 23$\pm 10$ pc
and an absolute magnitude of M$_V$ = -3.0$^{+2.0}_{-0.7}$.  One star
that is likely associated with this companion has an SDSS spectrum
confirming it as a blue horizontal branch star at 48 kpc.  The
color-magnitude diagram of SDSSJ1049+5103 contains few, if any,
horizontal or red giant branch stars, similar to the anomalously faint
globular cluster AM 4. The size and luminosity of SDSSJ1049+5103
places it at the intersection of the size-luminosity relationships
followed by known globular clusters and by Milky Way dwarf
spheroidals.  If SDSSJ1049+5103 is a globular cluster, then its
properties are consistent with the established trend that the largest
radius Galactic globular clusters are all in the outer halo.  However,
the five known globular clusters with similarly faint absolute
magnitudes all have half-mass radii that are smaller than
SDSSJ1049+5103 by a factor of $\gtrsim$ 5.  If it is a dwarf
spheroidal, then it is the faintest yet known by two orders of
magnitude, and is the first example of the ultra-faint dwarfs
predicted by some theories. The uncertain nature of this new system
underscores the sometimes ambiguous distinction between globular
clusters and dwarf spheroidals.  A simple friends-of-friends search
for similar blue, small scalesize star clusters detected all known
globulars and dwarfs closer than 50 kpc in the SDSS area, but yielded
no other candidates as robust as SDSSJ1049+5103.

\end{abstract}


\keywords{Milky Way: globular clusters ---
          galaxies: formation ---
          galaxies: dwarfs ---
          Local Group: surveys
          .}
\ifemulate\else
   \clearpage
\fi

\section{Introduction}
Milky Way globular clusters are invaluable pieces in the puzzle of
galaxy formation. At present, their properties support a general
picture of Galactic halo formation as a combination of accretion and
dissipative collapse (see review in \citealt{mackey04}). However, the
detailed interpretation of globular cluster (GC) properties in the
context of galaxy formation is complex. One outstanding problem is the
sometimes ambiguous distinction between GCs and dwarf spheroidal
galaxies (dSphs).  For example, a few Milky Way GCs, such as $\omega$
Cen, have a spread in stellar age and metallicity similar to that seen
in many dwarf galaxies \citep{ashman98}, and have absolute magnitudes
that overlap those of known dSph galaxies. A small number of faint GCs
have radial profiles that are well fit by an NFW profile
(e.g. Palomar 13;
\citealp{cote02}) or have central densities similar to those of dSphs
(e.g. Palomar 14; \citealp{harrisGCcat}) and thus may be the remnants
of a stripped dSph.

The relationship between globular clusters and dSphs is particularly
interesting in light of recent predictions for low mass substructure
around the Milky Way
(\citealp{klypin99,moore99,bullock00,benson02,susa04,kravtsov04},
among others).  It is difficult to determine whether GCs ever
contained a substantial amount of non-baryonic dark matter
\citep{ashman98}, which would arguably put them in the category of
ultra-faint dwarf galaxies.  If some globular clusters
are embedded in extended dark matter halos, the dark matter may not be
dynamically important within the extent of the observable stellar
distribution.


There are $\sim 150$ known globular clusters and 9 known dSphs
orbiting the Milky Way.  The total number of known clusters has 
increased by just a few percent over the last twenty-five years
\citep{harris97,harrisGCcat,ortolani00,hurt00,irwin95,ortolani93} and
nearly all of the new globular clusters lie at low Galactic latitude.
Only one Milky Way dSph has been discovered since 1990.  The lack of
new GCs or dSphs at $|b|$ $>$ 30$^{\circ}$ could lead some to believe
that all high latitude systems have been discovered.  However, one
anomalously faint GC (AM 4; M$_V$ = +0.2) was discovered serendipitously
more than 20 years ago
\citep{madore82}, suggesting that other ultra-faint star clusters
may still reside undetected in our halo.  Furthermore, the advent of
the Sloan Digital Sky Survey (SDSS; \citealt{york00}) could lead to
the discovery of similar systems, should they exist \citep{willman02}.
In this paper, we report the discovery of SDSSJ1049+5103, a new
ultra-faint, stellar system in the outer halo of the Milky Way. We
estimate and discuss some properties of SDSSJ1049+5103 in comparison
to both globular clusters and Milky Way dwarf spheroidal galaxies.

\section{Photometric Data}

\subsection{Sloan Digital Sky Survey and Object Discovery}

The Sloan Digital Sky Survey (SDSS; \citealt{york00}), is a
spectroscopic and photometric survey in 5 passbands ($u,g,r,i,z$;
\citealt{fukugita96,gunn98,hogg01,smith02}), that has thus far imaged
thousands of square degrees of the sky.  Data is reduced with an 
automatic pipeline consisting of: astrometry
\citep{pier03}; source identification, deblending and photometry
\citep{lupton01}; photometricity determination \citep{hogg01};
calibration \citep{fukugita96,smith02}; and spectroscopic data
processing \citep{edr}.

We discovered SDSSJ1049+5103
as part of an ongoing SDSS survey for Milky Way satellite galaxies
\citep{willman02}.  This object was detected at ($\alpha_{2000},
\delta_{2000}$) = (162.35,51.05) as a 12$\sigma$ fluctuation over the average spatially smoothed density of stellar sources with 21.0 $< r <$ 22.5.  See
\citealt{willman02} and Willman et al., 2004, in preparation, for details of
the survey analysis technique.  Although we analyzed $\sim 5000$
square degrees of available photometric data thus far, the data
relevant for this discovery are included in Data Release 2 of the SDSS
(DR2, \citealt{dr2}).

Figure~\ref{fig:wherryimage} is a 0.57 x 0.42 deg$^2$ $g,r,i$ image
centered on the detection.  Because SDSSJ1049+5103 is so sparse, it is
difficult to see in the image alone.  However, the stellar overdensity
is readily visible in the overplotted spatial distribution of faint
blue stars ($g - r$ $<$ 0.3). To more clearly illustrate the strength
of the overdensity, we show a spatially smoothed density map of stars
with $g-r < 0.65$ covering 0.5 $\times$ 0.5 degrees$^2$ around the
detection in Figure~\ref{fig:smoothimage}.  This figure shows that the
center of the cluster is detected at more than 20$\sigma$ over the
foreground when only blue stars are included in the analysis.  The
density contours do not exhibit obvious evidence for tidal stripping,
such as that seen around Palomar 5 \citep{rockosi02,odenkirchen03} as
well as numerous other Milky Way GCs \citep{leon00}.  However, a lack
of obvious tidal features in the SDSS data is unsurprising, because
the surface brightness of SDSSJ1059+5103 is so faint. Therefore,
deeper observations may reveal tidal distortion in the stellar
distribution.

\placefigure{fig:wherryimage}
\placefigure{fig:smoothimage}

Our algorithm for detecting satellite galaxies is not optimized for
the discovery of small scale length blue stellar overdensities, such
as SDSSJ1049+5103. Therefore, to investigate whether numerous such
systems remain undetected in the Milky Way's halo, we performed a
friends-of-friends search for groups of stars with $g - r < 0.3$ and
$r > 23$.  We used a linking length of 0.8$'$ and examined groups with
as few as 5 stars.  Although this simple search recovered both
SDSSJ1049+5103 and all of the known globular clusters and dSphs closer
than 50 kpc in the area searched, no obvious new candidates were
found.  Unfortunately, AM 4, the lowest luminosity of the known
clusters, is not in the SDSS area.  It is thus unclear whether a
comparably faint GC would have been detected with a simple
friends-of-friends approach.  Furthermore, the method we used is only
sensitive to very blue star clusters closer than $\sim$ 50 kpc.  It
was nevertheless surprising that there appeared to be no other systems
similar to SDSSJ1049+5103 in the $\sim$ 5000 deg$^2$ currently covered
in our search.  However, if the Milky Way GC luminosity function
(GCLF) at ultra-faint magnitudes does not deviate from that observed
between -4.0 $< M_V <$ -7.4, one would not expect to discover many
additional globular clusters.  Extrapolating the known GCLF
\citep{mclaughlin94,mclaughlin96} to faint magnitudes predicts a total
of only a few undiscovered GCs fainter than M$_V$ = -4.0 over the
whole sky.

\subsection{Follow-up Observations} \label{sec:apo}

On June 10, 2004, we obtained follow-up imaging of SDSSJ1049+5103 on
the 3.5-m telescope at Apache Point Observatory.  We used the SpiCAM
2048x2048 CCD, which has a resolution of 0.282 arcseconds
pixel$^{-1}$.  Three 900 second exposures and one 600 second exposure
were taken in the SDSS $g$ filter, and 1200, 900, and 600 second
exposures were taken in the SDSS $r$ filter.  Seeing was $\sim$ 1.6$'$
in $g$ and $1.4'$ in $r$ and observations were taken at high
airmass. These combined observations are thus only sufficient to
resolve stars as faint as $r$ $\sim$ 23.  The total sky coverage of
these data is $\sim$ 60 arcmin$^2$.  These data were photometrically
calibrated by comparison with SDSS observations of the same field.

Figure~\ref{fig:apoimage} is a 0.2 x 0.075 deg$^2$ $g,r$ image of the
APO data.  An overdensity of faint stars is visible near the center.

\placefigure{fig:apoimage}

\section{Results} \label{sec:results}

\subsection{Color-Magnitude Diagrams}

Figure~\ref{fig:sdsscmd} shows the color-magnitude diagram (CMD) of
SDSSJ1049+5103 and of the surrounding field as observed by SDSS.  The
stars in the 'source' CMDs include all those within the central
1.75$'$, which roughly corresponds to the half-light radius of the
source (see \S3.2).  The SDSS imaging data become incomplete near r
= 21.5, because star-galaxy separation is unreliable at fainter magnitudes
\citep{ivezic00}. These data have been corrected for reddening, using
the maps of \citet{schlegel98}.

\placefigure{fig:sdsscmd}

The CMD of SDSSJ1049+5103 contains an overabundance of stars bluer
than $g - r$ = 0.5 relative to the field.  We consider three broad
possibilities for the nature of these blue stars:

\begin{enumerate}
\item a young, metal rich stellar population with a main sequence turnoff around $g - r$ = 0.3 
\item an old, metal poor stellar population with a main sequence turnoff around $g - r$ = 0.3
\item a horizontal branch plus a few red giant branch stars
\end{enumerate}

Both a young, metal rich and an old, metal poor stellar population
could have a main sequence turnoff with $g - r \sim$ 0.3.  If the
stars in SDSSJ1049+5103 with $g - r \sim$ 0.3 are indeed main sequence
turnoff stars, then the stars with $g - r$ = 0.45 and 20 $< r <$ 21
are sub-giant branch stars.  However, those stars are bluer relative
to the detected turnoff than sub-giant stars of a young ($<$ 10 Gyr)
stellar population (see isochrones in
\citealt{girardi04}).  We therefore consider it unlikely that
SDSSJ1049+5103 is a young, metal rich stellar population.

To distinguish between the second and third possibilities, we compare
the CMD of SDSSJ1049+5103 to those of several low luminosity globular
clusters.  We compare to empirical rather than theoretical isochrones
because the Main Sequence colors of theoretical isochrones in Sloan
filters may be offset from those of actual old stellar populations
\citep{girardi04}.  Figure~\ref{fig:palcmds} shows the CMDs of Palomar
5, Palomar 15, and Palomar 3 as observed by SDSS, with the empirically
derived stellar locus of Pal 5 projected to the correct solar distance
and overplotted on each CMD.  The data in these plots have been
corrected for reddening using the maps of
\citet{schlegel98}.  The Pal 5 stellar locus does provide a reasonable
match to both Pal 3 and Pal 15's stars, but with a slight shift in
color due to metallicity differences.  Although Pal 5's stellar
population has been shown to display mass segregation
\citep{koch04}, it is nonetheless an acceptable basis for comparison
because it is the most nearby, and thus the most well measured of the
sparse globulars in the SDSS area.

We overplotted the stellar locus of Palomar 5 on the CMD of
SDSSJ1049+5103 in Figure~\ref{fig:sdsscmd}. Considering the
substantial photometric errors on stars fainter than $r$ = 21.5 in the
SDSS, the Pal 5 stellar locus projected to 45 kpc and to 170 kpc (plus
an offset in color) both provide reasonable matches to the data.  If
the nearby distance is correct, then the blue stars are turnoff stars.
If the far distance is correct, then they are horizontal branch stars.
The star at ($g - r$, $r$; $\alpha_{2000}, \delta_{2000}$) =
(-0.32,19.8; 162.3048, 51.0424) has an SDSS spectrum (plate-mjd-fiber
876-52669-375) that shows it is a blue horizontal branch star at a
distance of 48 kpc, supporting the hypothesis that SDSSJ1049+5103 is
an old stellar system near d = 45 kpc.

\placefigure{fig:palcmds}

The deeper CMD based on the APO data, shown in
Figure~\ref{fig:apocmd}, provides even more compelling evidence that
the detected stellar overdensity is a turnoff at 45 kpc rather than a
horizontal branch at 170 kpc.  Pal 5 has an age of 11 - 12 Gyr
\citep{martell02} and an [Fe/H] = -1.38 \citep{harrisGCcat}. The main sequence turnoff (MSTO) of
SDSSJ1049+5103 is bluer than that of Pal 5.  The bluer turnoff color
may mean that this new companion is more metal poor than Pal 5,
although the small number of resolved stars in the existing data makes
the metallicity difficult to estimate. We assign a generous
uncertainty of $\pm 10$ kpc to the distance estimate to account for
the fact that SDSSJ1049+5103's turnoff may be intrinsically more or
less luminous than that of Pal 5 (e.g. it would be intrinsically
brighter if its stars are more metal poor and of a similar age as Pal
5's).

A few blue straggler candidates are visible in the CMD bluer than $g -
r = 0.15$ and brighter than $r = 21.5$. Assuming that we are seeing
the turnoff of an old, metal-poor population, SDSSJ1049+5103 contains
very few stars brighter than the sub-giant branch.  One known globular
cluster, AM 4, also appears to be devoid of any horizontal branch or
red giant branch stars.  In \S3.3 we evaluate the significance of the
dearth of evolved stars in SDSSJ1049+5103.

\placefigure{fig:apocmd}

 Figure~\ref{fig:gcdistr} shows the X,Z distribution of the known Milky Way
globulars and dSphs with the new detection overplotted. Our estimated distance
of 45 kpc from the Sun places SDSSJ1049+5103 at 50 kpc from the center
of the Galaxy.  If SDSSJ1049+5103 is indeed a globular cluster, it
will add to the small number of globulars known to have Galactocentric
distances greater than 35 kpc.

\placefigure{fig:gcdistr}

\subsection{Radial Profile}
Figure~\ref{fig:profile} shows the azimuthally averaged radial profile
of SDSSJ1049+5103.  Because our APO observations do not have
sufficient area to properly subtract the foreground, we used a cut of
$g - r < 0.65$ and $r < 22.5$ to eliminate the majority of foreground
field stars from both the APO and the SDSS observations.
Figure~\ref{fig:profile} shows that the SDSS stars satisfying these
criteria approach a field density of $\sim 0.22$ stars arcmin$^{-1}$
by 4.5 arcmin from the detection center.  The dashed line denotes this
adopted foreground level.  The profile is consistent with a power law,
with a possible break near 2$'$, and shows no evidence for a core at
the center.  However, the central radial bin in this plot has a radius
of 1.0$'$ (13 pc at a distance of 45 kpc), so any core would likely be
unresolved by the current data.  The small number of stars also
prevents us from measuring a reliable central surface brightness.

We corrected the radial stellar counts for the foreground level
overplotted on Figure~\ref{fig:profile} and estimated the half-light
radius from the resulting cumulative radial profile shown in
Figure~\ref{fig:cumfraction}.  This estimate assumes that the stellar
population is roughly constant with radius.  The half-light radius,
$r_{1/2}$, is a good way to characterize the initial size of stellar
systems, because it changes slowly with their dynamical evolution
(\citealp{murphy90}, among others).  Both the SDSS data and the APO
data yield $r_{1/2} \sim 1.75'$, which corresponds to a physical size
of 23 pc at a distance of 45 kpc.  Allowing for a generous uncertainty
in $r_{1/2}$ of $\pm 0.5'$ and including a distance uncertainty of
$\pm$ 10 kpc, we estimate a plausible range of physical half-mass
radii of 13 to 36 pc.

If it is a globular cluster, then SDSSJ1049+5103 follows the well
known trend that all large size GCs are in the outer Galactic halo
\citep{vdbergh03}.  Pal 14 is the only known GC with a half-mass radius larger than 20 pc.

\placefigure{fig:cumfraction}

\subsection{Stellar Luminosity Function and Total Luminosity}
We use three approaches to estimate the total luminosity of
SDSSJ1049+5103.  We first estimate a lower limit by summing the
luminosity of likely cluster stars within the half-light radius, and
then doubling the summed luminosity to account for stars outside the
half-light radius.  Taking all stars with $g - r < 0.65$ and $20.3 <
r < 23.0$, and accounting for the liberal distance uncertainty stated
above, this approach yields M$_{V,faint}$ = -1.5 $\pm$ 0.5.

Second, we compare the observed stellar luminosity function of the new
object to that of Palomar 5.  Table 1 shows the stellar luminosity
functions of SDSSJ1049+5103, as observed by both SDSS and APO, and of
Pal 5 projected to 45 kpc.  We include all stars bluer than $g
- r$ = 0.65 in the luminosity function of SDSSJ1049+5103.  The sharp
increase at faint magnitudes in the ratio of SDSSJ1049+5103 stars
observed at APO to Pal 5 stars observed in SDSS is due to the fact
that SDSS does not resolve stars as faint as the APO observation.
The numbers in this Table show that SDSSJ1049+5103 has $\lesssim$ 1/5
of the number of Pal 5 stars in each of the magnitude bins bright
enough to be well resolved by SDSS.  We thus divided Pal 5's
luminosity by the conservatively small factor of 5 to yield
$M_{V,bright}$ = -3.3. However, Table 1 shows that SDSSJ1049+5103 has
few, if any, stars brighter than $r
\sim$ 20.5, which means that it has few, if any, horizontal branch or
red giant branch stars.  We thus crudely correct $M_{V,bright}$ for
the fact that $\sim 30\%$ of Pal 5's luminosity comes from stars
brighter than the sub-giant branch and find M$_{V,corr}$ = -3.0, which
we adopt as the absolute magnitude of SDSSJ1049+5103 for the rest of
this paper.  Accounting for distance uncertainty, we derive a
maximum plausible luminosity of $M_V = -3.7$ with this technique,
resulting in a total range of -1 $< M_V <$ -3.7.

Similar to SDSSJ1049+5103, the globular cluster AM 4 has no stars
brighter than its main sequence turnoff. By comparison with M3, 
\citet{inman87} estimated that AM 4 should have 9 $\pm$ 1 stars
brighter than its turnoff; however it only has 1.  SDSSJ1049+5103 is
not as anomalous as AM 4 in that respect.  By comparison with Pal 5,
we estimate that there should be $\sim$ 7 stars in SDSSJ1049+5103 with
an apparent magnitude brighter than 20.5. The APO observations contain
4 candidates for such stars: ($g - r,r$) = (0.41,20.12; -0.33,19.81 -
the BHB star; 0.60,19.52; 0.56,18.17).  It is plausible that the
dearth of bright, red stars in SDSSJ1049+5103 is simply due to its low
stellar surface density.  Furthermore, Pal 5 has been shown to exhibit
radial mass segregation.  This segregation causes stars at the bright
end of Pal 5's luminosity function to be overrepresented in its
central region, relative to what one would expect for an unrelaxed
system. This bias could result in an overestimate of the expected
number of horizontal and red giant branch stars for SDSSJ1049+5103.

\section{Comparison With the Properties of Known Stellar Systems}

We now compare the properties estimated above to those of known
globular clusters and dSphs.

\subsection{M$_V$ and $r_{1/2}$}
We compare the estimated half-light radius, $r_{1/2}$, and absolute
magnitude of SDSSJ1049+5103 to those of the known Milky Way globular
clusters and the Milky Way dSphs (except for Sagittarius), in
Figure~\ref{fig:GCdsphprops}.  We estimate the half-light radii of the
dSphs using data from \citet{mateo98} to determine the geometric mean
of each core and tidal radius along the semi-major and semi-minor axes
and then integrating the corresponding King model. We shaded the
empirical size-luminosity locus followed by both the globulars and the
dSphs.  Because there are so few known MW dSphs, their locus is not
robustly known.  We thus overplotted the red galaxies from the SDSS
low luminosity galaxy catalog of \citet{blanton04}.  The Milky Way
dSphs follow nearly the same size-luminosity relation followed by
other red, low luminosity galaxies.

\placefigure{fig:GCdsphprops}

SDSSJ1049+5103's combination of size and luminosity places it at the
intersection of the relationships followed by globular clusters and
and by the nearby dSphs.  Although SDSSJ1049+5103 is 6 magnitudes
fainter than the faintest known Milky Way dwarf, its low surface
brightness re-raises the timely question: ``What is the difference
between globular clusters and dwarf galaxies?''  The presence of dark
matter is the apparent physical, and perhaps the fundamental,
distinction between the two sets of objects.  The fact that globular
clusters are much more compact than dwarfs is the most easily measured
and most reliable observational criterion for classification.  However
Figure~\ref{fig:GCdsphprops} shows that the size-luminosity
relationships of globular clusters and of Milky Way dSphs overlap at
low luminosities, highlighting the vague distinction between these two
classes of objects. Furthermore, the 6 magnitudes separating the
faintest Milky Way dwarfs and SDSSJ1049+5103 have not yet been
uniformly searched for dwarfs.  New surveys may uncover additional
nearby faint galaxies, and then SDSSJ1049+5103 would not be such an
outlier from other dwarfs.

Indeed, \citet{benson02} predict the existence of Milky Way dwarf
satellite galaxies as faint as the faintest GCs and with half mass
radii that roughly follow the same luminosity-size relation as the
known dSphs.  Three known GCs also fall within the overlapping
size-luminosity region: AM 1, Pal 5, and Pal 14.  Pal 5 is well known
to currently be undergoing massive disruption by the Milky Way
\citep{rockosi02,odenkirchen03}.  Pal 14 is a young globular cluster known to have the lowest central concentration of any known GC, and AM 1 is the most distant Milky Way GC (d = 120 kpc; \citealt{harrisGCcat}).  Pal 14 and AM 1 are obvious candidates to search for dark matter in nearby globular clusters. 

Figure~\ref{fig:GCdsphprops} also shows that SDSSJ1049+5103 is more
than $\gtrsim5\times$ the physical size of other faint globulars.
However, the fact that SDSSJ1049+5103 is an apparent outlier in size
from other faint GCs could be due to observational bias.  A globular
cluster with a larger scale size than a cluster of the same total
luminosity is more difficult to detect than the more compact cluster.
Furthermore, all known large scale length GCs are in the outer halo
\citep{vdbergh03}. It is thus possible that other ultra-faint, large
scale size GCs exist, but have not yet been detectable because they
lie at outer halo distances where far fewer of their stars are
resolved than if they were more nearby.  The lack of other candidates
identified by out friends-of-friends search may argue against this
possibility.

\subsection{Mass}
If SDSSJ1049+5103 is a globular cluster, Figure 10 suggests that it
has an anomalously large half-light radius.  This raises the
possibility that it is a globular cluster undergoing tidal disruption.
In this section, we do a crude calculation of cluster mass and tidal
radius to investigate whether the present data are consistent with this
interpretation.

\citet{mandushev91} used the dynamical masses of 32 globular clusters to derive an empirical relationship between cluster mass and absolute magnitude:

\begin{equation}
log(M_{\mathrm GC}/M_{\odot}) = -0.456M_V + 1.64
\end{equation}

\noindent Given the result from \S3.3 that $-3.7 < M_V < -1.0$, this equation
yields 10$^{2.1} M_{\odot} < M < 10^{3.3} M_{\odot}$ for
SDSSJ1049+5103.  We note that all of the clusters used in their study
were brighter than M$_V$ = -5.6, so the reliability of the
extrapolation to M$_V$ $\sim$ -3.0 is quite uncertain.

We estimate the tidal radii corresponding to this range of
satellite masses using the equation:

\begin{equation}
r_{\mathrm tidal} \sim R_{GC} (\frac{\mathrm M_{GC}}{3M_{\mathrm MW}})^{1/3}
\end{equation}

\noindent from \citet{binney87}.  In this equation, $R$ is the satellite's Galactocentric
distance, and $M_{MW}$ is the total mass of the Milky Way within that
distance.  We calculated r$_{tidal}$ assuming R$_{GC}$ = 50 kpc and
that $v_c$ = 220 km/sec at that distance (as recently shown to be the
case by \citealt{bellazzini04}).  The mass range estimated above
yields 22 pc $< r_{tidal} < $ 40 pc (1.7$'$ -- 3.0$'$ at a solar
distance of 45 kpc).  {This range of tidal radii is an upper limit
on $r_{\mathrm tidal}$ for SDSSJ1049+5103, if it is truly a low
mass-to-light system such as a globular cluster. The satellite's
pericentric distance could be much smaller than its present
Galactocentric distance, which would result in smaller derived tidal
radii.  Its radial profile does not exhibit a break until $r \sim
3.5'$, and it reaches the foreground stellar density at $r \sim 6'$.
If the tidal radius of SDSSJ1049+5103 is actually $\lesssim 1.7 -
3.0'$, one may expect the stellar profile to exhibit a break
characteristic of tidal stripping at $r < 3.5'$.  However, the data
are not yet deep enough to produce a robust measurement of the stellar
distribution. The existing data thus allow the possibilities that
SDSSJ1049+5103 is a low mass-to-light system that might be tidally
stripped, or that the stars are embedded in a more extended, higher
mass-to-light system.

\section{Summary and Future Work}
In this paper, we report the discovery of SDSSJ1049+5103, a new
stellar system that is likely in the outer halo of the Milky Way.
Based on comparison with Palomar 5, this new system appears to be
$\sim$ 50 kpc from the Galactic center, have a half light radius of 23
pc, and M$_V = -3.0^{+2.0}_{-0.7}$.  SDSSJ1049+5103 has a size and
luminosity that places it at the intersection of the size-luminosity
locus followed by Milky Way globular clusters and and that followed by
Milky Way dSphs and nearby faint red galaxies.  Both the fundamentally
ambiguous distinction between some globular clusters and dSph galaxies
and the fact that SDSSJ1049+5103 is unusual relative to the vast
majority of GCs leaves open the possibility that it is an extreme
dwarf galaxy nearly two orders of magnitude fainter than Ursa Minor,
the faintest known MW dSph.  Furthermore, some theories predict the
presence of low central surface density, ultra-faint dSphs
(e.g. \citealp{benson02}) such as SDSSJ1049+5103.  If SDSSJ1049+5103
is a globular cluster, then its properties are consistent with
undergoing tidal disruption.

Neither the \citet{willman02} survey nor a friends-of-friends search
revealed additional companions similar to, or even a bit fainter than,
SDSSJ1049+5103 in the 5000 square degrees analyzed thus far.  This
suggests that there is not a substantial unknown population of similar
companions closer than $\sim$ 50 kpc.

We are in the process of obtaining both deep, wide field imaging to
accurately measure the spatial distribution of SDSSJ1049+5103 and
spectra of individual stars to measure ages, metallicities, and
line-of-sight velocities.  Specifically, deeper imaging may
distinguish between a King and NFW surface brightness profile and may
also reveal tidal features, which would provide strong constraints on
its current mass \citep{moore96}.


\acknowledgements
    BW and JJD were partially supported by the Alfred P. Sloan
    Foundation.  MRB, DWH, and NW were partially supported by NASA
    (grant NAG5-11669) and NSF (grants PHY-0101738 and
    AST-0428465). DPS acknowleges support from NSF grant AST03-07582.
    We thank David Martinez-Delgado, David Schlegel, and Douglas
    Finkbeiner for useful discussions and data analysis software.  We
    also thank the anonymous referee for comments that improved this
    paper.

  The SDSS is managed by the Astrophysical Research Consortium (ARC) for
  the Participating Institutions. The Participating Institutions are The
  University of Chicago, The U.S. Department of Energy's Fermi National
  Accelerator Laboratory, The Institute for Advanced Study, The Japan
  Participation Group, The Johns Hopkins University, The Korean
  Scientist Group, Los Alamos National Laboratory, the
  Max-Planck-Institute for Astronomy (MPIA), the Max-Planck-Institute
  for Astrophysics (MPA), New Mexico State University, University of
  Pittsburgh, Princeton University, the United States Naval Observatory
  and the University of Washington.
  
  Funding for the project has been provided by the Alfred P. Sloan
  Foundation, the Participating Institutions, the National Aeronautics
  and Space Administration, the National Science Foundation, the
  U.S. Department of Energy, the Japanese Monbukagakusho, and the Max
  Planck Society.


\ifsubmode\else
\baselineskip=10pt
\fi


\clearpage


\clearpage

\begin{figure}
\plotone{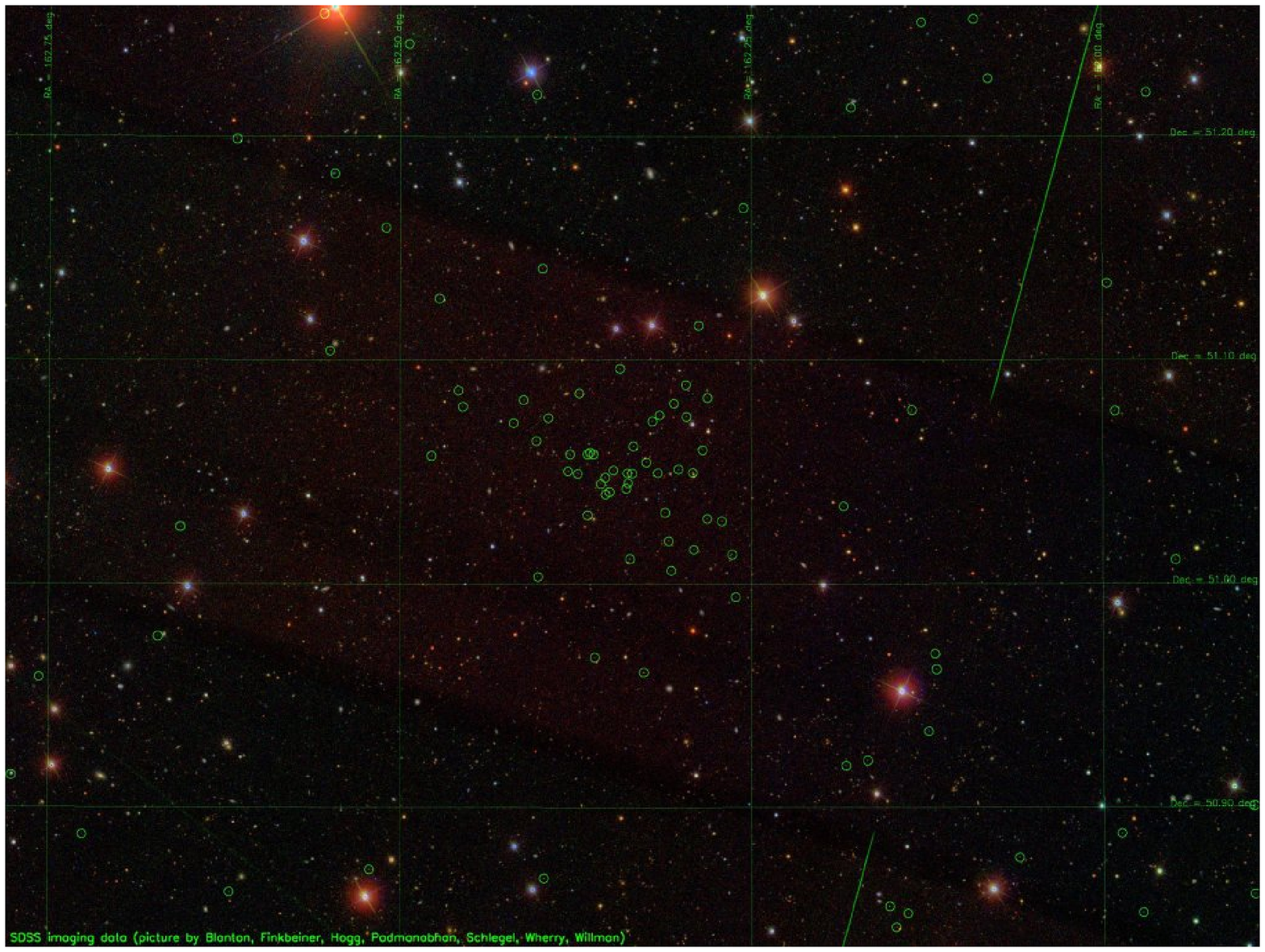}
\caption{SDSS true color $g,r,i$ image of 0.57 x 0.42 deg$^2$ centered on the detection.  Stellar sources with colors consistent with blue horizontal branch and main sequence turnoff stars ($g - r <$ 0.3) are circled in green.  The image is made with color-preserving nonlinear stretches \citep{lupton04}.}
\label{fig:wherryimage} 
\end{figure}

\clearpage

\begin{figure}
\plotone{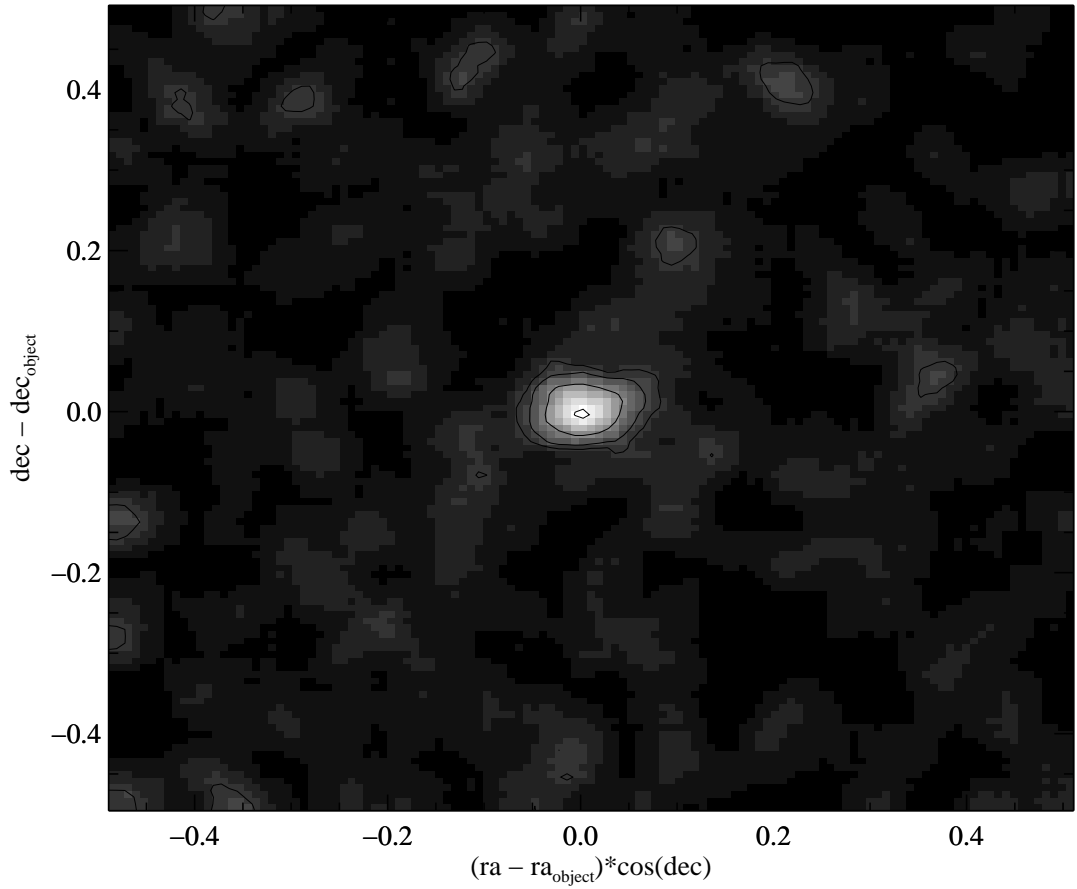}
\caption{Smoothed image of stars with $g - r <$ 0.65 and in a 0.5 x 0.5 deg$^2$ field centered on the detection.  The contours represent smoothed stellar densities of 3, 5, 10, and 20$\sigma$ above the foreground.}
\label{fig:smoothimage} 
\end{figure}

\clearpage

\begin{figure}
\plotone{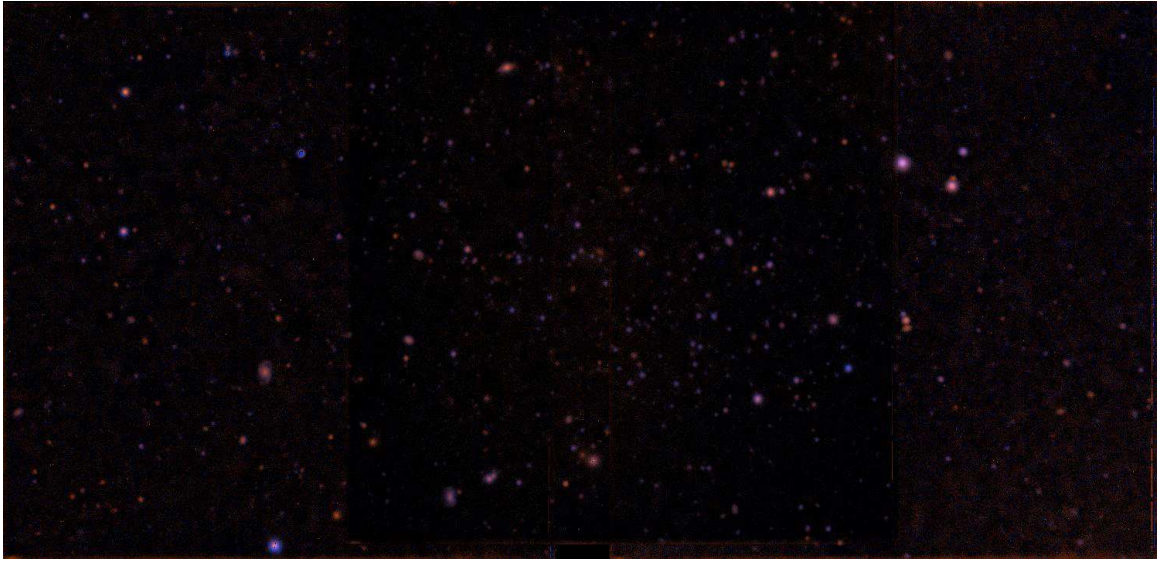}
\caption{APO true color $g,r$ image of 0.2 x 0.075 deg$^2$ centered on the detection.}
\label{fig:apoimage} 
\end{figure}

\clearpage

\begin{figure}
\plotone{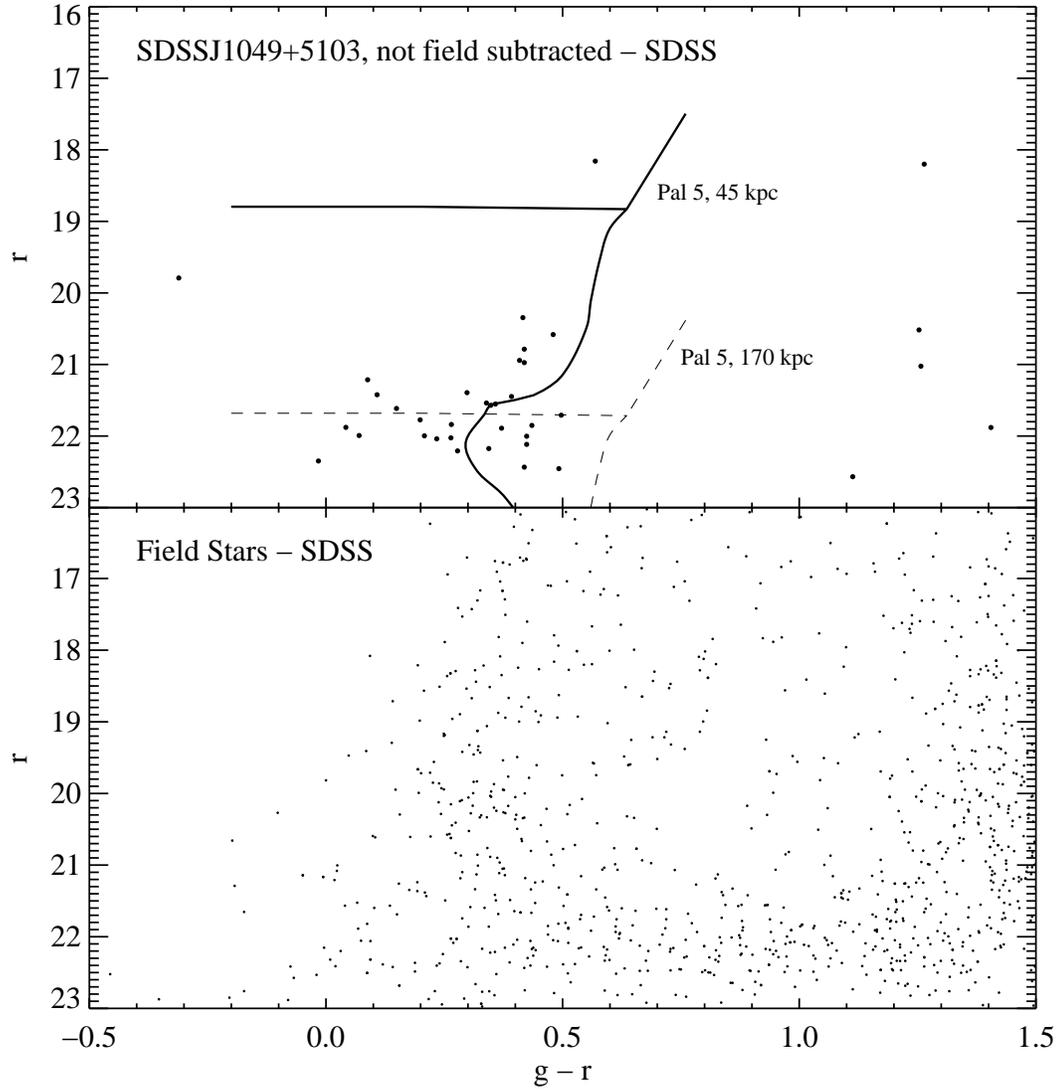}
\caption{Color-magnitude diagram of the source and of the surrounding field stars, as observed by SDSS.  The source CMD includes all stars within 1.75$'$ of the center, and has not been field subtracted. The field CMD includes all stars within 0.5 degrees of the center.  The stellar locus of Pal 5 stars that we empirically measured with SDSS data and projected to 45 and 170 kpc, is overplotted.  These data have been corrected for reddening \citep{schlegel98}.}
\label{fig:sdsscmd} 
\end{figure}

\clearpage

\begin{figure}
\plotone{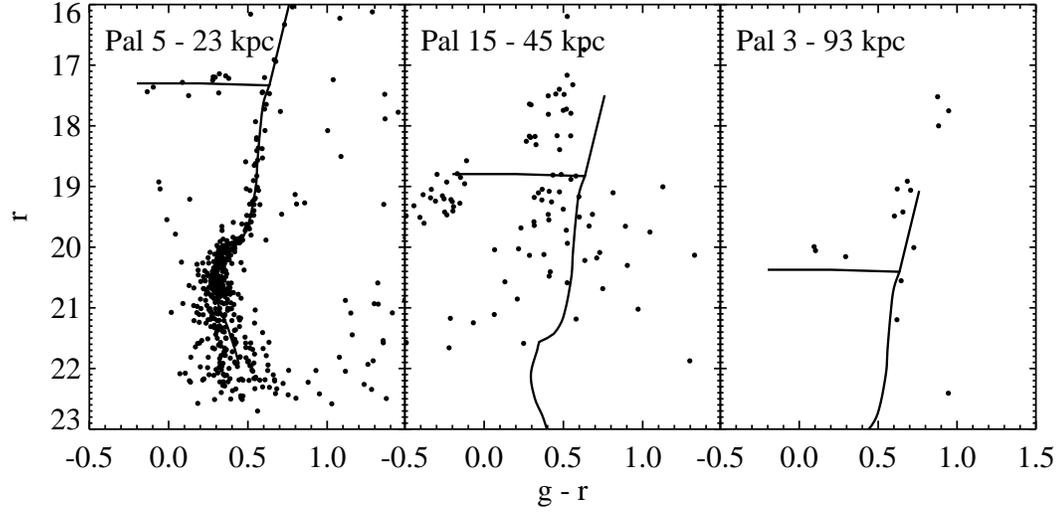}
\caption{Color-magnitude diagrams of the known globular clusters Pal 5, Pal 15, and Pal 3.  All stars within their published half-mass radii \citep{harrisGCcat} are included on the CMDs.  Pal 5's empirically derived stellar locus is projected to the distance of each cluster and overplotted for reference.   These data have been corrected for reddening \citep{schlegel98}. Cluster distances are from \citet{harrisGCcat}. }
\label{fig:palcmds} 
\end{figure}

\clearpage

\begin{figure}
\plotone{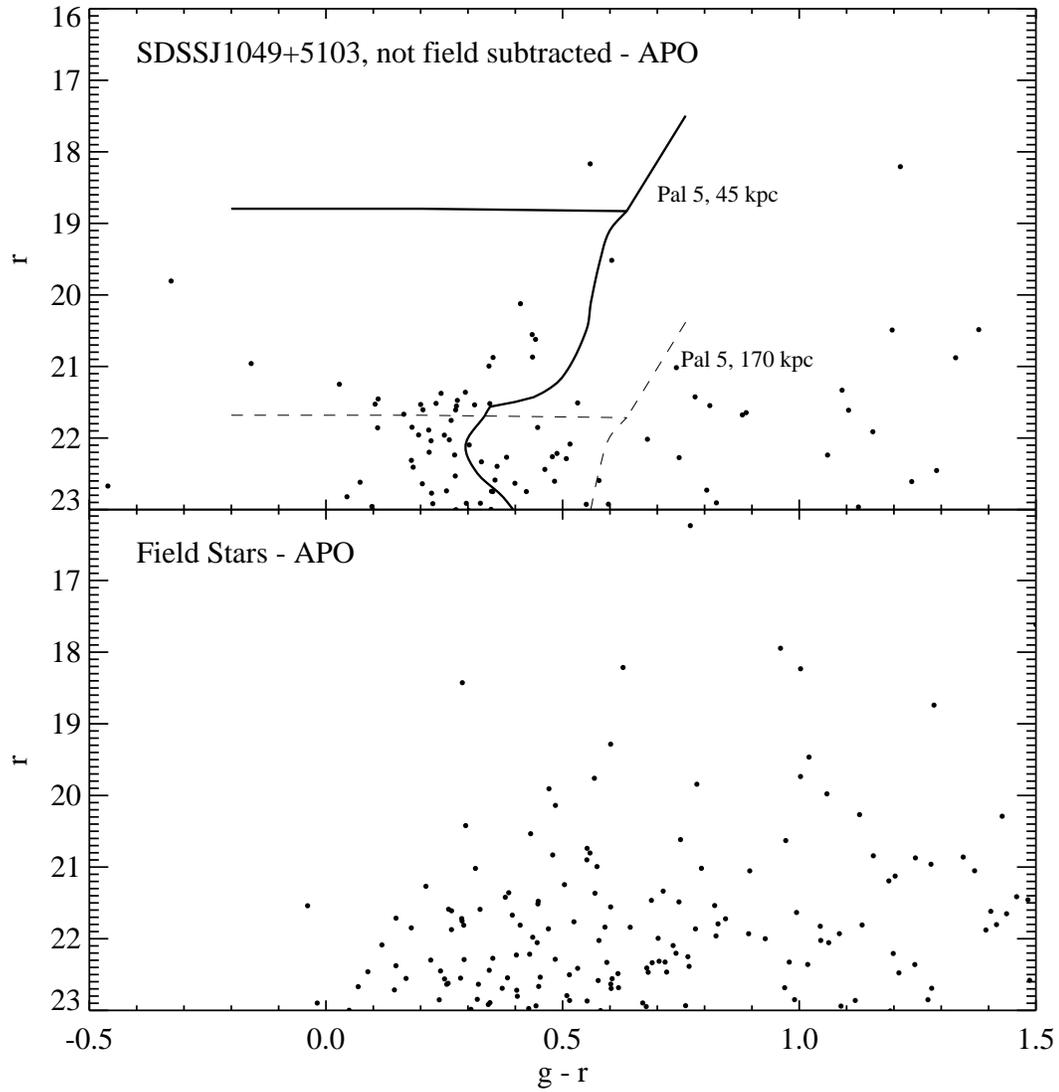}
\caption{Color-magnitude diagram of the source and of the surrounding field stars, as observed in follow-up observations at APO.  As in Figure~\ref{fig:sdsscmd}, the source CMD includes all stars within 1.75$'$ of the center, and is not field subtracted. Seeing was $\sim$ 1.6$'$
in $g$ and $1.4'$ in $r$ and observations were taken at high
airmass. The field CMD includes all other stars in the entire $\sim$60
arcmin$^2$ follow-up area. The stellar locus of Pal 5 stars,
empirically measured with SDSS data and projected to 45 and 170 kpc,
is overplotted. These data have been corrected for reddening
\citep{schlegel98}.}
\label{fig:apocmd} 
\end{figure}

\clearpage

\begin{figure}
\plotone{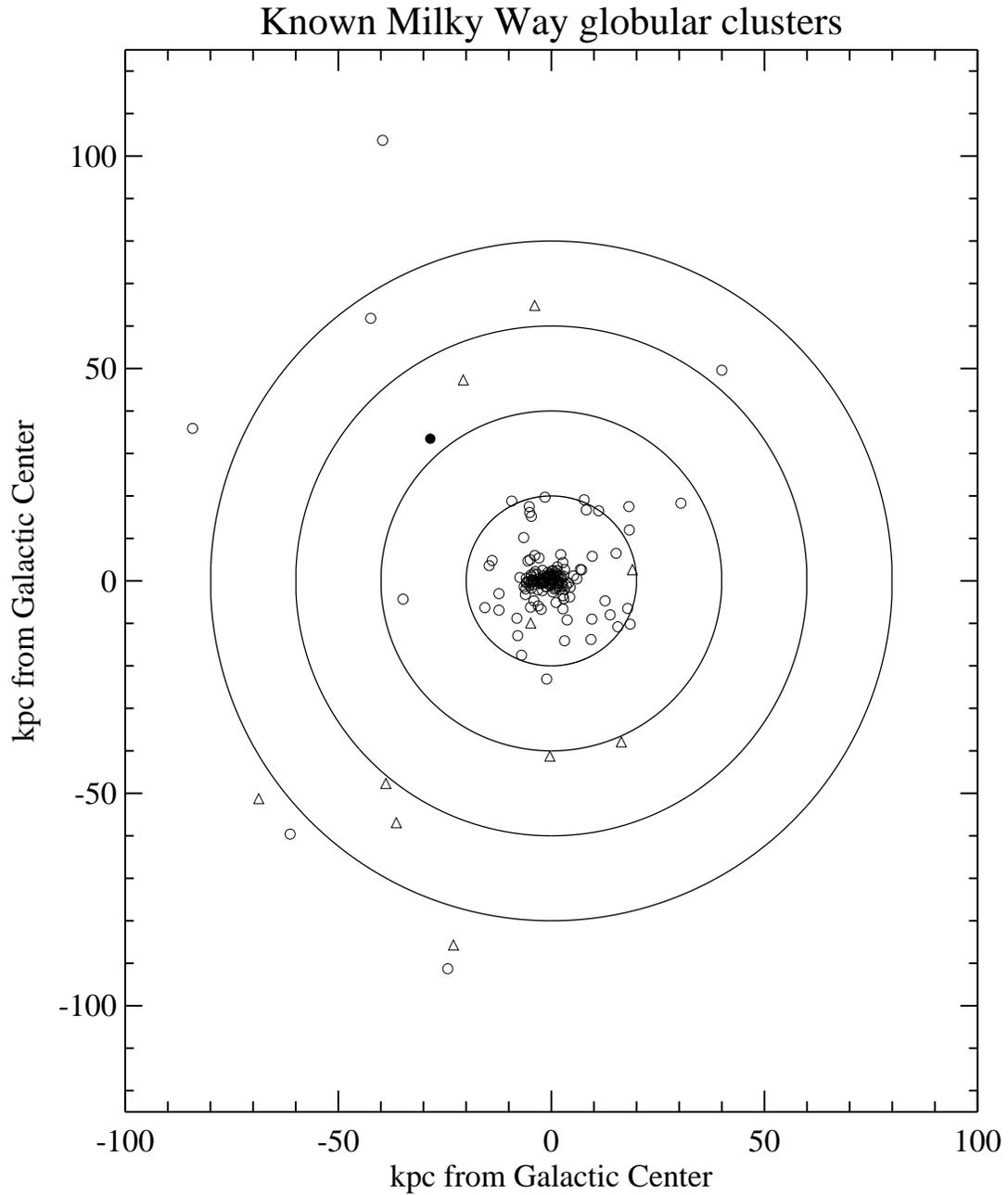}
\caption{Spatial distribution of the (X,Z) positions of 150 known globular clusters (open circles; \citealt{harrisGCcat}), of 10 known nearby dwarfs (open triangles; \citealt{mateo98}), and of the new companion (filled circle).  The large circles show projected Galactocentric distances of 20, 40, 60, and 80 kpc.  The Galactic disk is oriented perpendicular to the y-axis.}
\label{fig:gcdistr} 
\end{figure}

\clearpage

\begin{figure}
\plotone{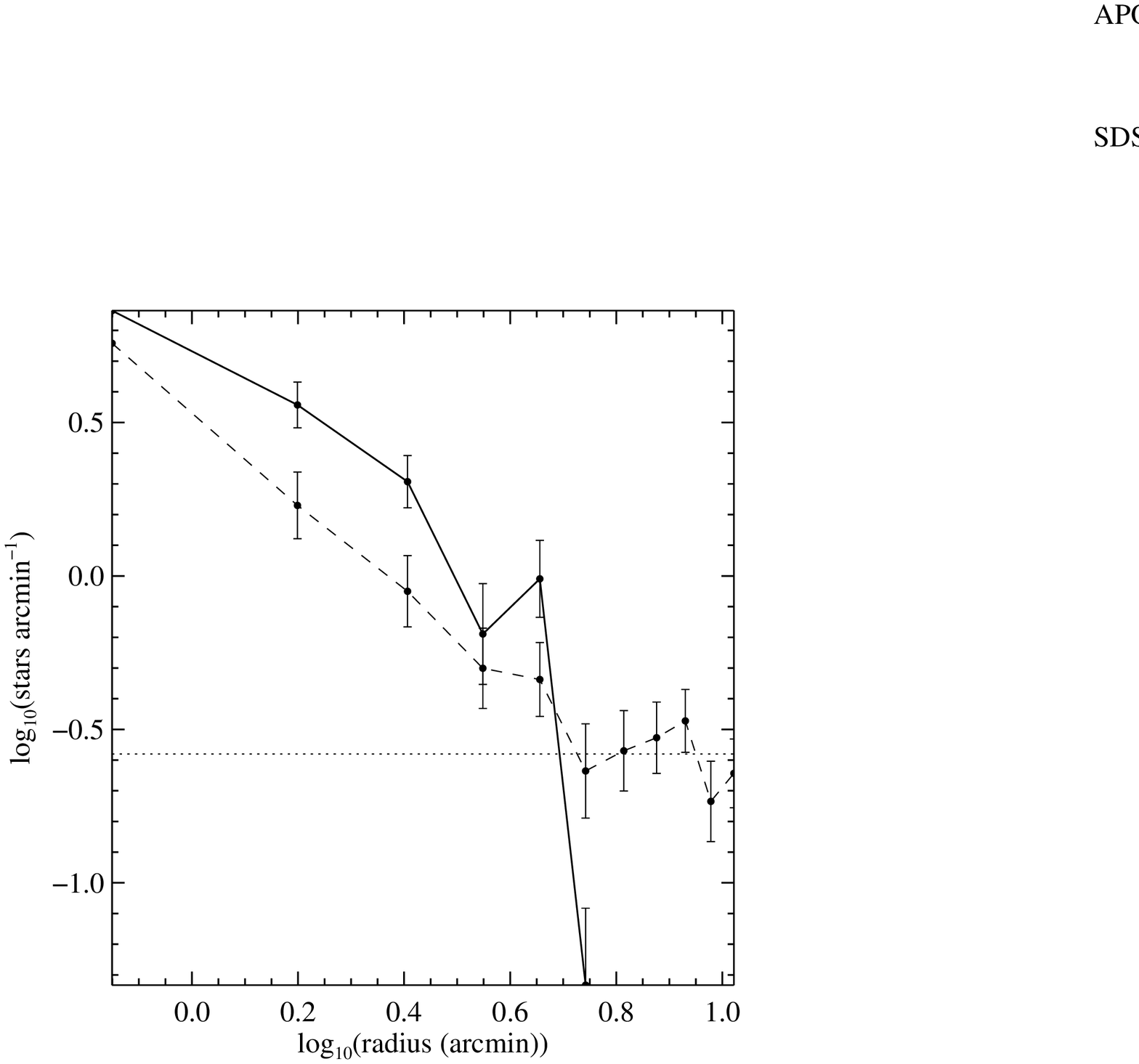}
\caption{Radial profile of the stellar number density observed for the detection.  The dotted line shows the adopted foreground stellar density of stars bluer than $g - r$ = 0.65 and brighter than $r$ = 22.5.  Error bars were calculated assuming Poisson statistics.}
\label{fig:profile} 
\end{figure}

\clearpage

\begin{figure}
\plotone{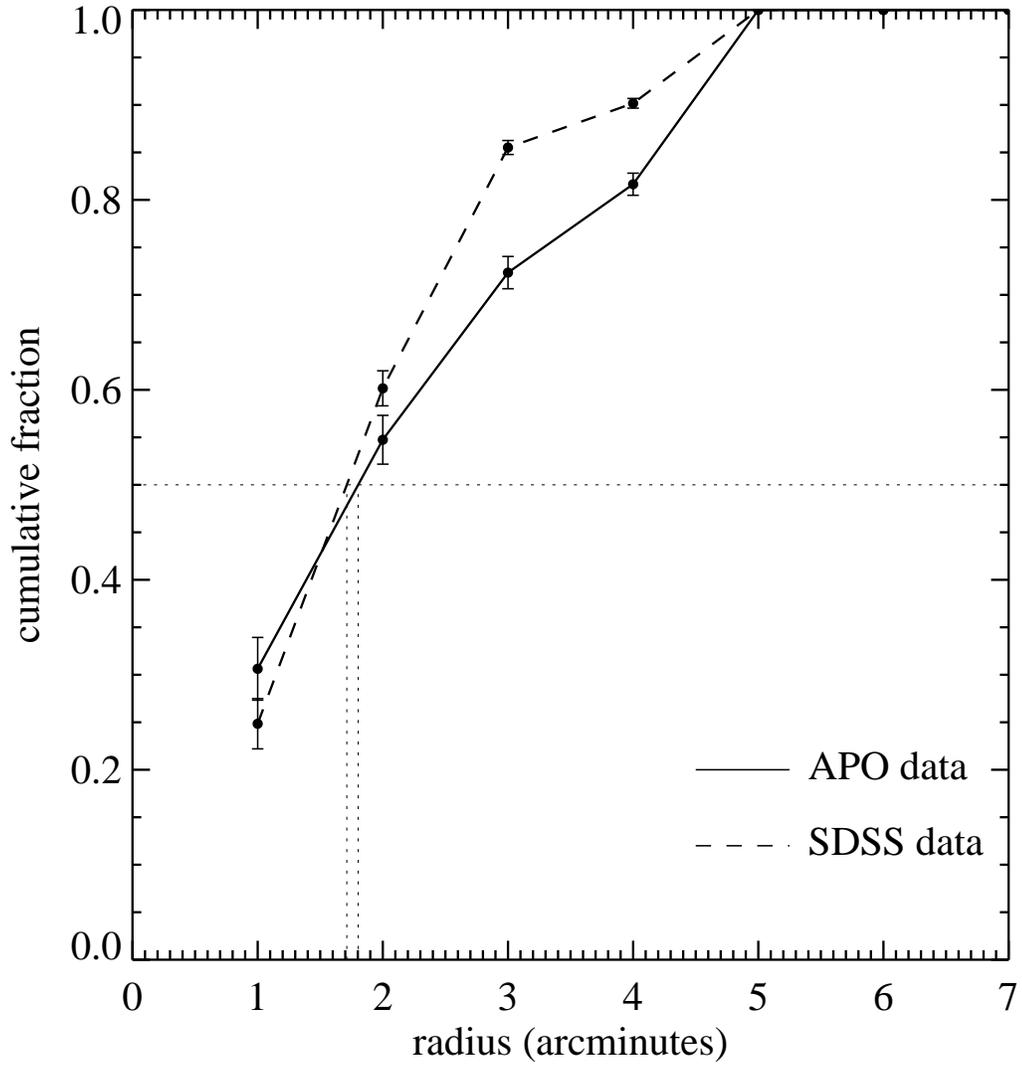}
\caption{Cumulative radial distribution of stars in the detection that are bluer than $g - r$ = 0.65 and brighter than $r$ = 22.5.  The cumulative fraction is corrected for the foreground stellar density overplotted in Figure~\ref{fig:profile}, and forced to be 1.0 at the radius beyond which observed stellar density reaches the foreground level.  The dotted lines show the half-mass radii found with the SDSS and the APO data, assuming a constant stellar population with radius.  Error bars were calculated assuming Poisson statistics.}
\label{fig:cumfraction} 
\end{figure}

\clearpage

\begin{figure}
\plotone{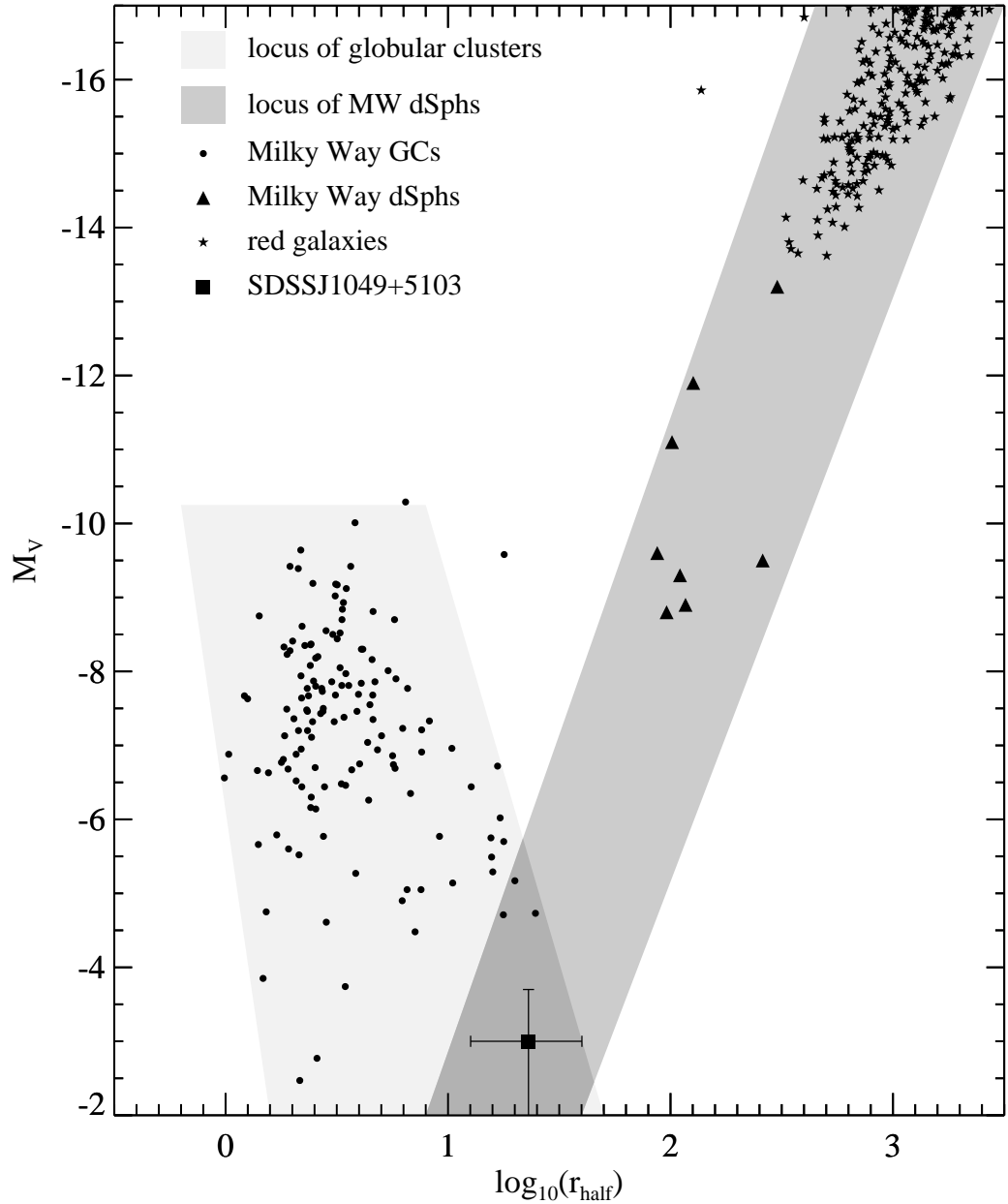}
\caption{The absolute magnitudes and half-light radii of Milky Way globular clusters (circles), dwarf spheroidal galaxies (triangles), faint red galaxies in the SDSS (stars; \citealp{blanton04}), and SDSSJ1049+5103 (square).  AM 4 is too faint (M$_V$ = +0.2) to be included on this plot. The approximate loci of the globular cluster and the dwarf spheroidal data are shaded.  The Milky Way dSphs appear to follow a similar size-luminosity relation as other faint red galaxies. Data is from \citet{harrisGCcat} and \citet{mateo98}.}
\label{fig:GCdsphprops} 
\end{figure}


\clearpage


\ifsubmode\pagestyle{empty}\fi

\begin{table}
\begin{tabular}{cccc}
\hline
m$_r$ & N(SDSS)\tablenotemark{a,b} & N(APO)\tablenotemark{a,b} &
N(Pal5)\tablenotemark{a,c} \\ \hline $<$ 20.0 & 1 & 3 & 29 \\
\smallskip 20.0 - 20.5 & 1 & 1 & 8 \\ \smallskip 20.5 - 21.0 & 4 & 6
& 20 \\ \smallskip 21.0 - 21.5 & 4 & 5 & 53 \\ \smallskip 21.5 - 22.0
& 12 & 17 & 103 \\ \smallskip 22.0 - 22.5 & 9 & 16 & 114 \\ \smallskip
22.5 - 23.0 & 0 & 21 & 78 \\ \smallskip 23.0 - 23.5 & 0 & 31 & 81 \\
\smallskip 23.5 - 24.0 & 0 & 18 & 57 \tablenotetext{a}{The number of
stars within the half-light radius, 1.75$'$ for SDSSJ1049+5103 and
2.9$'$ for Palomar 5, from \citet{harrisGCcat}.}
\tablenotetext{b}{These numbers only include stars that are bluer than
$g-r = 0.65$.}  \tablenotetext{c}{These numbers have been properly
corrected for field stars, and the luminosity function has been
projected to 45 kpc.}  \end{tabular} \tablecaption{Stellar Luminosity
Function of SDSSJ1049+5103} \end{table} 
\end{document}